\documentclass[twocolumn,prb,aps,showpacs,groupedaddress]{revtex4-1}

\usepackage{graphicx}
\usepackage{amsmath}
\usepackage{float}

\begin{document}

\title{Impact of the skyrmion spin texture on magnetoresistance}
\author{Andr\'{e}~Kubetzka}
\author{Christian Hanneken}
\author{Roland Wiesendanger}
\author{Kirsten von Bergmann}
\email{kbergman@physnet.uni-hamburg.de}

\affiliation{Department of Physics, University of Hamburg, D-20355 Hamburg, Germany}

\date{\today}

\begin{abstract}

We investigate the impact of the local spin texture on the differential conductance by scanning tunneling microscopy. In the focus is the previously found non-collinear magnetoresistance, which originates from spin mixing effects upon electron hopping between adjacent sites with canted magnetic moments. In the present work it is studied with lateral resolution both for the zero magnetic field spin spiral state as well as for individual magnetic skyrmions at different magnetic field values. We analyze in detail the response of the differential conductance and find different dependencies of peak energy and peak intensity on the local properties of the non-collinear spin texture. We find that in the center of a skyrmion the peak energy and intensity scale roughly linear with the angle between nearest neighbor moments. Elsewhere in the skyrmion, where the non-collinearity is not isotropic and the magnetization quantization axis varies, the behavior of the peak energy is more complex.

\end{abstract}

\maketitle

\section{Introduction}

When magnetic materials are involved in transport measurements the spin has an impact on the resistance, in general coined magnetoresistance (MR). The most prominent example is the giant magnetoresistance (GMR)~\cite{Baibich_PhysRevLett.61.2472,Binasch_PhysRevB.39.4828}, which occurs when current flows between two magnetic electrodes separated by a non-magnetic metallic spacer. A similar setup is used to measure tunneling magnetoresistance (TMR)~\cite{Julliere1975}, where the metallic spacer is exchanged with an insulating tunnel barrier. Because in both cases the conductance depends on the relative alignment of the magnetization in the two electrodes, one of them can be used to read out the magnetic state of the other one. The TMR can be exploited in a spatially resolved scanning tunneling microscopy (STM) measurement, when the tip is spin-polarized; such spin-polarized STM measurements can reveal spin structures at surfaces down to the atomic scale~\cite{Wiesendanger2009,Bergmann2014}. Changes in MR can be described in scattering models~\cite{Levy_PhysRevLett.79.5110} or based on electronic band structure effects~\cite{Bode2002}. Because in scanning tunnel spectroscopy (STS) the differential conductance (d$I$/d$U$) reflects the vacuum local density of states (LDOS), STM can be used to study the latter. Recently, the MR change due to the presence of a domain wall has been discussed~\cite{Levy_PhysRevLett.79.5110,Marrows_PhysRevLett.92.097206}. However, different types of MR are expected in the presence of a domain wall, and a separation of different effects is experimentally difficult because the measured signal is a combination of several effects. Due to its spatial resolution STM has been a key method to unravel the changes of the vacuum LDOS of a domain wall: the difference between the center of the wall (in-plane magnetization) and a domain (out-of-plane) was detected with a non-magnetic tip and it was pinned down by accompanying density functional theory (DFT) calculations of the electronic band structure to a mixing of states due to spin-orbit coupling (SOC), resulting in tunnel anisotropic magnetoresistance (TAMR)~\cite{Bode2002}.

Recently, STM measurements have identified an MR effect, which originates from the non-collinearity of a spin texture~\cite{Hanneken_NatNano2015}. The observed non-collinear magnetoresistance (NCMR), see sketch in inset to Fig.\,\ref{fig1}(a), arises due to spin mixing effects because of electron hopping between atoms with canted magnetic moments, as demonstrated by two-band tight-binding as well as ab-initio DFT calculations. These theoretical approaches for the system of a PdFe bilayer on Ir(111) have demonstrated that the electronic band structure of this material is different for collinear versus non-collinear spin configurations~\cite{Hanneken_NatNano2015}. This explains the experimental observation that skyrmions are electronically different compared to their ferromagnetic surrounding, enabling an all-electrical detection scheme.

NCMR is a general effect, but it is not always straightforward to disentangle from other MR effects, in particular the TAMR, which typically leads to a maximal MR change for perpendicular magnetization quantization axes. While the TAMR originates from SOC and is thus usually limited to a few percent, the NCMR results from a spin mixing and values up to $50\%$ have been found experimentally for PdFe~\cite{Hanneken_NatNano2015}, with MR defined by:
\begin{equation}
	 \mathrm{MR} = \frac{(\mathrm{d}I / \mathrm{d}U)_\mathrm{FM} -  (\mathrm{d}I / \mathrm{d}U)_\mathrm{NC}} {(\mathrm{d}I / \mathrm{d}U)_\mathrm{FM}} \times 100\%,
\end{equation}
where d$I$/d$U$ is the differential conductance signal at a non-collinear (NC) or ferromagnetic (FM) position of the spin texture. Furthermore, whereas the TAMR is an on-site effect, which depends on the local quantization axis of the magnetization, NCMR is governed by the details of the spin texture in the local environment.

A theoretical investigation including SOC of a closely related PdFe system, i.e.\ with a different stacking of the Pd layer, has evaluated the contributions of non-collinearity and SOC to the energy-dependent MR in the skyrmion center: although both contributions are comparable for extremely small skyrmions, the SOC has a minor effect for more realistically sized skyrmions, however, the effect of the non-collinearity was found to be as large as $20\%$~\cite{Crum_NatComm_2015}. Other theoretical studies of spin spirals in atomic chains~\cite{Czerner_PSSB:PSSB201046190} and films~\cite{Seemann_PhysRevLett.108.077201} have also found that the impact of SOC on the total MR is less than $10\%$, and that the conductance change due to the non-collinearity is dominating the transport. Recent experimental studies have shown that discrete jumps in magnetic field dependent MR can be linked to the appearance and disappearance of individual skyrmion tubes in a wire~\cite{Du_NatCom2015}.

In this work we extend the investigation of the previously studied system of PdFe on Ir(111) and present a more detailed analysis, in particular of the central finding of the response of a susceptible peak in the differential conductance and its relation to the non-collinearity~\cite{Hanneken_NatNano2015}. First, we present data which demonstrates that the NCMR effect occurs also for the zero magnetic field spin spiral ground state and in both stackings of the Pd on Fe. Then we turn to the magnetic skyrmions, that arise in this system upon application of a perpendicular magnetic field~\cite{Romming.2013}. We look closely into the magnetic field dependent details of the skyrmion spin texture~\cite{Romming_PhysRevLett.114.177203} and correlate the evolution of the peak energy as well as its intensity with the different parameters that characterize the local spin configuration. We conclude with a discussion about the different local contributions to the total MR in these non-collinear spin textures.

\section{Experimental details}

The experiments were performed in a multi-chamber ultra-high vacuum system with separate chambers for substrate cleaning, metal deposition, and STM measurements. The Ir crystal was cleaned by repeated cycles of sputtering and annealing, and from time to time heating in oxygen. Iron was deposited onto the Ir(111) surface at elevated temperature to obtain step-flow growth of the fcc stacking~\cite{Bergmann2006} and subsequently Pd was deposited~\cite{Romming.2013}. Two different STMs were used with base temperatures of 4.2\,K and 8\,K, and the tip material was either Cr or W. Maps of differential conductance (d$I$/d$U$) were measured simultaneously to constant-current ($I$) topographic images at identical sample bias voltage ($U$) with closed feedback loop. Scanning tunnel spectroscopy (STS), d$I$/d$U$($U$), was performed with open feedback loop; the stabilization parameters ($U_{\rm stab}$ and $I_{\rm stab}$) before switching off the feedback loop were chosen with particular care to ensure that the tip-sample distance does not depend on the magnetic state of the sample, i.e.\ in first approximation the laterally-resolved spectroscopic data represents constant-height spectroscopy. The d$I$/d$U$ signal was measured via lock-in technique where the bias voltage was modulated with $U_{\rm mod}$ at a frequency of about 2.6\,kHz.

\section{Experimental results}

\begin{figure}[]
\includegraphics[width=1\columnwidth]{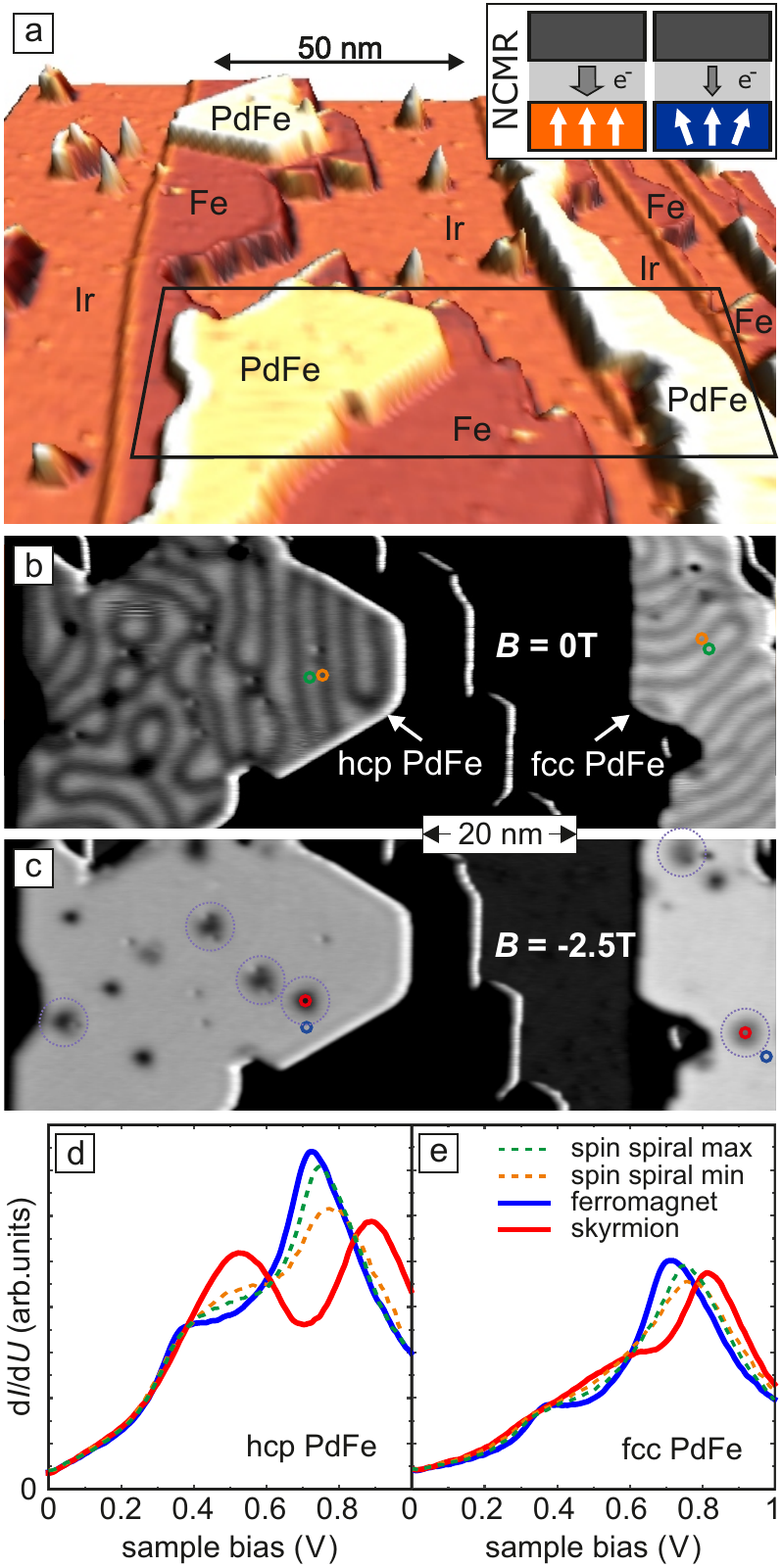}
\caption{\textbf{(a)}~Pseudo-3D STM topography of a typical sample of Pd and Fe on Ir(111), colorized with the differential conductance signal; the inset is a sketch of the NCMR effect for a planar tunnel junction: the resistance/conductance between a non-magnetic electrode and a magnetic electrode is different for a collinear compared to a non-collinear spin texture. \textbf{(b)}~d$I$/d$U$~map of the area indicated by the black rectangle in (a) showing NCMR contrast for the spin spiral ground state in both stackings of the Pd on Fe. \textbf{(c)}~d$I$/d$U$~map of the same area as in (b) in applied magnetic field with individual magnetic skyrmions imaged by NCMR. \textbf{(d,e)}~d$I$/d$U$~spectra of the different magnetic states as color-coded in (b,c) for both stackings of PdFe. Measurement parameters for (a-c): $U=+0.7$\,V, $I=1$\,nA, $U_{\rm mod}=40$\,mV; for (d,e) $U_{\rm stab}=-1$\,V, $I_{\rm stab}=1$\,nA, $U_{\rm mod}=7$\,mV; all $T=4.2$\,K.}
\label{fig1}
\end{figure}

The system of the atomic bilayer PdFe on Ir(111) serves as our model system and Fig.\,\ref{fig1}(a) shows an overview STM constant-current image, colorized with the simultaneously obtained d$I$/d$U$ signal. The measurement was performed with a non-magnetic tip and the dark irregular stripes connected to the step edges of the Ir(111) substrate indicate the Fe monolayer grown in fcc stacking. The islands on top of the Fe are monolayer Pd patches and different signal strengths indicate different stackings. A d$I$/d$U$ map of the area indicated by the black rectangle is shown again in Fig.\,\ref{fig1}(b) with an hcp PdFe island on the left and an fcc PdFe stripe to the right. In addition to the different contrast levels due to the Pd stacking one can observe a stripe pattern indicative of the spin spiral states: the observed d$I$/d$U$ contrast originates from the NCMR at this sample bias of +0.7\,V. Because of the non-negligible out-of-plane anisotropy of PdFe~\cite{Dupe2014,Romming_PhysRevLett.114.177203} the spin spiral is inhomogeneous and the d$I$/d$U$ signal is higher for out-of-plane regions with smaller angles between neighboring magnetic moments as compared to in-plane magnetic moments with larger angles between them~\cite{Hanneken_NatNano2015}. Note that while there is no external magnetic field applied this is not the virgin state and some interconnected in-plane rings have remained after a previous magnetic field sweep.

When an external magnetic out-of-plane field is applied single magnetic skyrmions can be found in PdFe of both stackings, see Fig.\,\ref{fig1}(c) for a d$I$/d$U$ map of the same area at $B=-2.5$\,T. At this magnetic field strength the skyrmions appear as dark spots due to the NCMR (see dotted circles); the skyrmions form due to the interface-induced Dzyaloshinskii-Moriya interaction and thus have unique rotational sense and are of cycloidal nature~\cite{Romming.2013,Dupe2014,Bergmann2014}. To study the electronic states of PdFe in the different magnetic phases local STS has been employed: Fig.\,\ref{fig1}(d) and (e) show spectra at positions of highest and lowest d$I$/d$U$ signal of the spin spiral (compare colored circles in (b)) as well as for the ferromagnetic region and the skyrmion center (compare colored circles in (c)) for hcp and fcc PdFe, respectively. All spectra are featureless at negative sample bias (not shown) but show characteristic peaks in the unoccupied states: the ferromagnetic states have a single peak at around $+0.7$\,V, which shifts towards higher energy for the spin spiral states and the skyrmion centers; at the same time the intensity of this peak decreases. As demonstrated previously~\cite{Hanneken_NatNano2015} a two peak structure is found for the skyrmion center of the hcp PdFe, and this change of the electronic structure has been modeled by tight-binding based on density function theory (DFT) calculations for the ferromagnetic state which neglect SOC~\cite{Hanneken_NatNano2015} and has been shown to originate from NCMR. In addition, the vacuum LDOS, which correlates with the experimentally measured d$I$/d$U$, has been obtained from these DFT calculations for the ferromagnetic state and a spin spiral with a period of 5.14\,nm, very similar to the experimentally observed one. For both stackings the agreement between the calculated vacuum LDOS and the experimentally obtained spectra is very nice, except for a rigid shift by about 0.2\,V~\cite{Hanneken_NatNano2015,SHBD2017}. Other DFT calculations for extremely small skyrmions for fcc PdFe, this time including SOC, have been published in Ref.\,\onlinecite{Crum_NatComm_2015} and show a peak at about 0.6\,V for the ferromagnetic state, which shifts to smaller energies upon introducing non-collinearity, in contrast to the experimentally observed shift of the peak at 0.7\,V to higher energies.

\begin{figure}[]
\includegraphics[width=1\columnwidth]{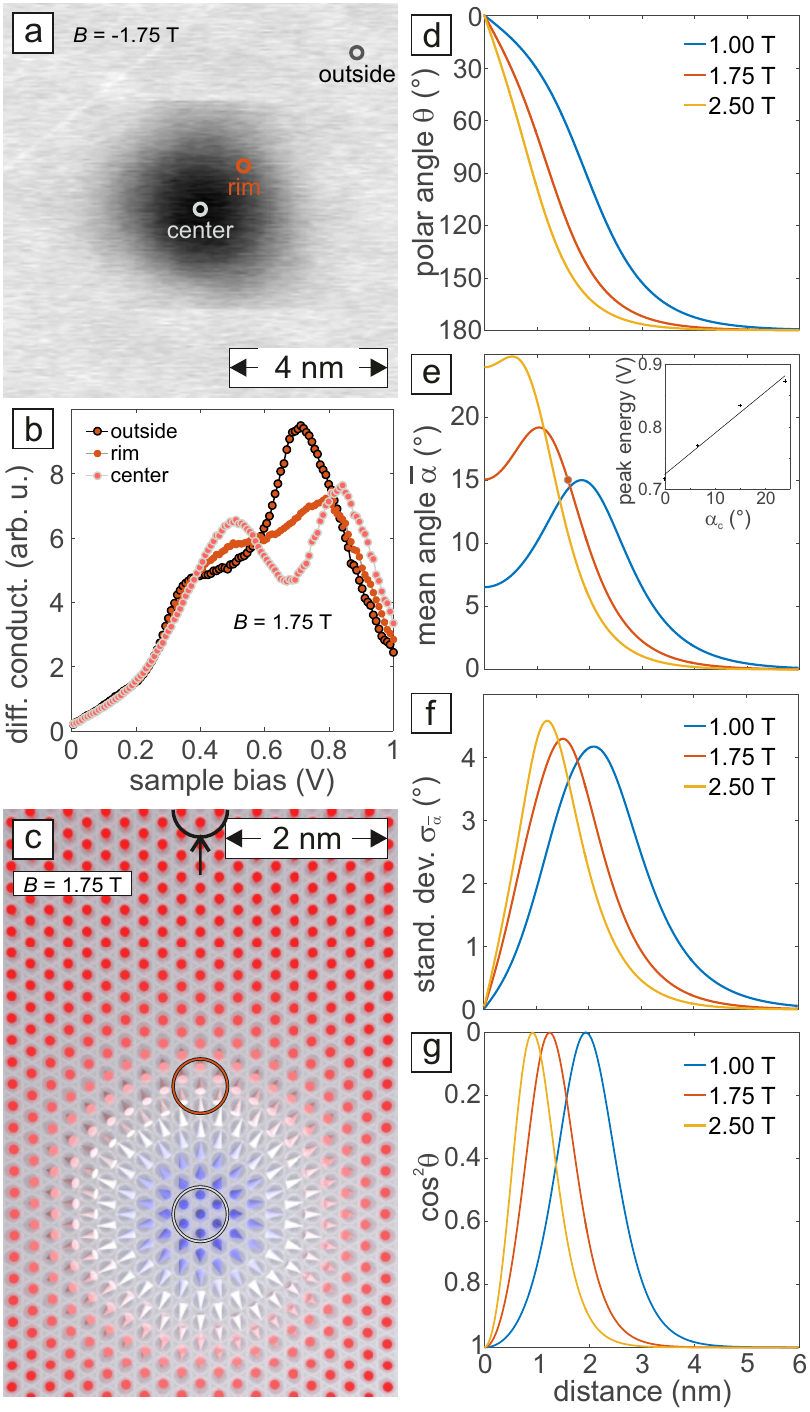}
\caption{\textbf{(a)}~d$I$/d$U$~map of a single skyrmion in PdFe on Ir(111) at $B=-1.75$\,T measured with a W tip ($U=+0.7$\,V, $I=1$\,nA, $U_{\rm mod}=20$\,mV, $T=8$\,K). \textbf{(b)}~d$I$/d$U$~spectra taken at different positions within the skyrmion (see color-coded circles in (a); $U_{\rm stab}=-0.3$\,V, $I_{\rm stab}=0.2$\,nA, $U_{\rm mod}=20$\,mV, $T=8$\,K). \textbf{(c)}~Sketch of a skyrmion at 1.75 T, the cones represent atomic magnetic moments, their directions point in the direction of magnetization, and the color-scale indicates the out-of-plane magnetization component (red down, blue up). \textbf{(d)-(g)}~Magnetic field dependent properties of individual magnetic skyrmions in PdFe as a function of distance from the skyrmion center: (d)~the polar angle of the spins in a skyrmion; (e)~the mean angle between a spin and its six neighbors and (f)~standard deviation of this mean angle; (g)~the magnetic quantization axis varies from out-of-plane in the center of a skyrmion via in-plane back to out-of-plane in its surrounding.}
\label{fig2}
\end{figure}

The question arises how the spectra spatially evolve between the skyrmion center and the ferromagnetic environment and if their characteristics can be directly correlated with the non-collinearity. A high resolution d$I$/d$U$~map of a single skyrmion in the hcp PdFe at $B=-1.75$\,T measured with a non-magnetic W tip is shown in Fig.\,\ref{fig2}(a). Spectra taken at different positions within this skyrmion are displayed in Fig.\,\ref{fig2}(b), see color coding in (a) for the positions with respect to the skyrmion center. From previous SP-STM studies~\cite{Romming_PhysRevLett.114.177203} the details of the spin texture of magnetic skyrmions in the hcp PdFe bilayer as a function of external magnetic field are known and Fig.\,\ref{fig2}(c) displays a sketch of the atomic magnetic moments within a magnetic skyrmion for an external magnetic field of 1.75\,T. The circles indicate the positions where the spectra displayed in Fig.\,\ref{fig2}(b) were obtained. The evolution of the size and shape of magnetic skyrmions in PdFe as a function of external magnetic field can be characterized by the polar angle $\theta$ of magnetization, as displayed in Fig.\,\ref{fig2}(d) for selected magnetic fields as a function of distance from the skyrmion center. This plot demonstrates that the diameter (defined as distance from in-plane to in-plane) changes from about 4\,nm at 1\,T to about 2\,nm at 2.5\,T, and that also the general shape of the skyrmion changes with external magnetic field.

Previous analysis~\cite{Hanneken_NatNano2015} has revealed that the peak energy in the center of a skyrmion scales roughly linear with the nearest-neighbor angle $\alpha_c$, see inset to Fig.\,\ref{fig2}(e). While for a spin in the center of a skyrmion the angle $\alpha$ to all six nearest-neighbors on the hexagonal lattice is the same ($\alpha_c$), i.e.\ the environment is symmetric, the situation is different for other positions within a skyrmion, compare configurations indicated by circles in the sketch of the spin structure of a magnetic skyrmion at 1.75\,T in Fig.\,\ref{fig2}(c). To account also for such inhomogeneous environments with different nearest-neighbor angles $\alpha$, we calculate the mean angle $\overline{\alpha}$ to the six nearest neighbors from the given spin texture and plot it as a function of distance from the skyrmion center, see Fig.\,\ref{fig2}(e). This parameter $\overline{\alpha}$ can be interpreted as the generalized degree of non-collinearity, and we find that its maximum value moves towards the center of the skyrmion as the external magnetic field is increased. The question arises, whether the peak energy directly reflects the local mean angle $\overline{\alpha}$, as suggested by the direct correlation of peak energy and nearest-neighbor angle $\alpha_c$ in the center of the skyrmion, see inset to Fig.\,\ref{fig2}(e). We can crosscheck this for the 1.75\,T skyrmion: both in the center of the skyrmion as well as at a distance of about 1.6\,nm from the center $\overline{\alpha}$ is about $15^{\circ}$. However, comparison of the experimental spectra obtained at the respective positions within the skyrmion (Fig.\,\ref{fig2}(b) center and rim) reveals that such a strict correlation between peak energy and mean angle $\overline{\alpha}$ cannot be confirmed.

Although the two selected positions within the 1.75\,T skyrmion (compare groups of atoms indicated by the circles in Fig.\,\ref{fig2}(c)) have the same mean angle $\overline{\alpha}$ there are also two obvious differences: first, the variation in the six nearest-neighbor angles $\alpha$, which can be quantified by the standard deviation $\sigma_{\overline{\alpha}}$ of the mean angle $\overline{\alpha}$, is different (compare also distance-dependent plots for selected magnetic fields in Fig.\,\ref{fig2}(f)). Naturally the standard deviation $\sigma_{\overline{\alpha}}$ is zero for the collinear state outside the skyrmion as well as for the symmetric spin arrangement in the center of the skyrmion. However, at all other positions within the skyrmion it is finite. The second apparent difference for the two positions with the same mean angle is the quantization axis of the magnetization, which is reflected by $\gamma = \mathrm{cos}^2(\theta)$ as plotted in Fig.\,\ref{fig2}(g): the magnetization quantization axis can also modify electronic states due to SOC, resulting in TAMR~\cite{Bode2002,Bergmann2012_PhysRevB.86.134422}. In a magnetic skyrmion a potential TAMR contribution $\gamma$ is identical in its center and outside, and it is maximal different for the in-plane spins, cf.\ Fig.\,\ref{fig2}(g). Because for the spectra displayed in Fig.\,\ref{fig2}(b) in the center of the skyrmion with $\gamma = 1$ and near the in-plane region with $\gamma = 0$ there is also a change in $\sigma$ from $0^{\circ}$ to about $4^{\circ}$, we cannot unambiguously identify the origin of this change, i.e.\ whether it is due to the change in $\sigma$ or $\gamma = 0$. However, experimentally we find a significant difference in the spectra in the center and outside the skyrmion, where $\sigma = 0^{\circ}$ and $\gamma = 1$ in both cases, which thus originates only from the change in the non-collinearity $\overline{\alpha}$.

\begin{figure}[]
\includegraphics[width=1\columnwidth]{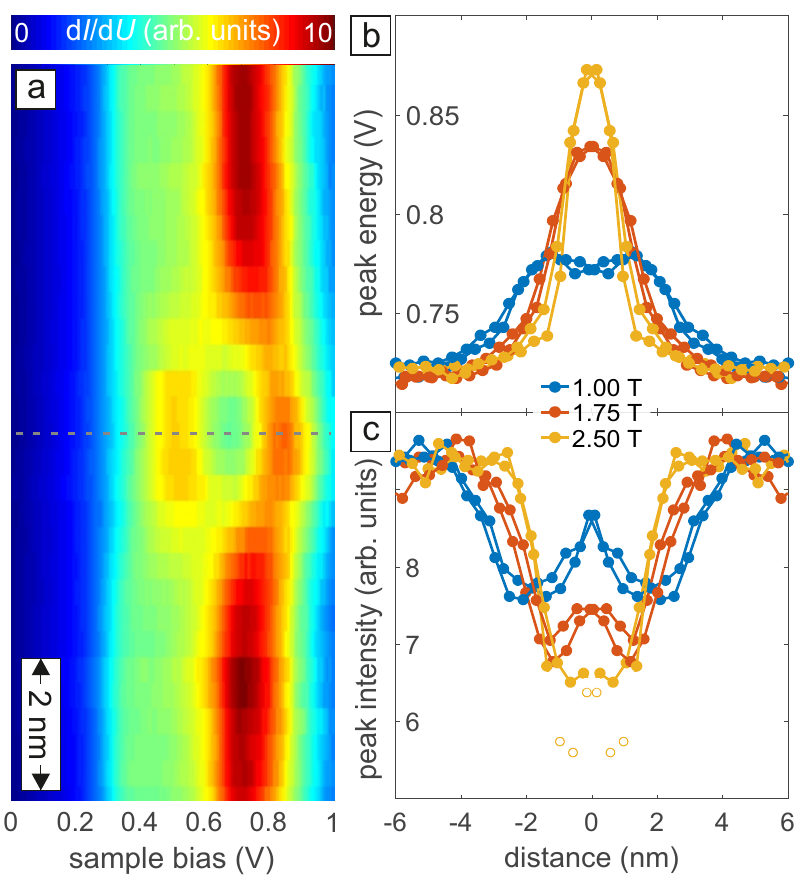}
\caption{\textbf{(a)}~Spatially-resolved d$I$/d$U$~spectra across a skyrmion at $B=-1.75$\,T measured with a W tip ($U_{\rm stab}=-0.3$\,V, $I_{\rm stab}=0.2$\,nA, $U_{\rm mod}=20$\,mV, $T=8$\,K); the dashed line indicates the skyrmion center. \textbf{(b)}~Evolution of the peak energy as a function of distance from the skyrmion center for different magnetic fields as labeled with the same W tip; all data points are also mirrored at the skyrmion center, lines are guides to the eye. \textbf{(c)}~Same as (b) but for the peak intensity; the empty circles at the bottom indicate data points that are considered unreliable due to a defect.}
\label{fig3}
\end{figure}

For a more detailed analysis of the changes of the electronic states due to the spin texture we measure the laterally-resolved spectra across a skyrmion with the same tip and identical parameters at different magnetic field values. Figure\,\ref{fig3}(a) shows a waterfall plot of the respective data set for a skyrmion at $B=-1.75$\,T, where each horizonal line is a single d$I$/d$U$ spectrum and the color code indicates the intensity; the center of the skyrmion is indicated by the dashed line. We find that the peak at around $+0.7$\,V in the ferromagnetic region outside the skyrmion shifts towards higher energy in the center of the skyrmion and at the same time changes its intensity. A plot of the evolution of peak energy and intensity for three different magnetic field values as a function of distance to the skyrmion center is shown in Fig.\,\ref{fig3}(b) and (c); the same data points are plotted both in one direction across the skyrmion and vice versa, i.e.\ they are mirrored at the center of the skyrmion; the lines are guides to the eye. For small skyrmions the maximum peak energy is in the center of the skyrmion, whereas for a large skyrmion at $B=-1$\,T a slight reduction of the peak energy is found near the center. The peak intensities, Fig.\,\ref{fig3}(c), show a local maximum in the center of the skyrmion for both $-1$\,T and $-1.75$\,T (not the entire data set measured at $-2.5$\,T can be analyzed quantitatively due to a defect near the center of the skyrmion which seems to have an impact on the intensity, and the unreliable data points are plotted as empty circles). In the following we try to correlate this magnetic field dependent evolution of peak energy and intensity with the details of the local spin texture.

\begin{figure}[]
\includegraphics[width=1\columnwidth]{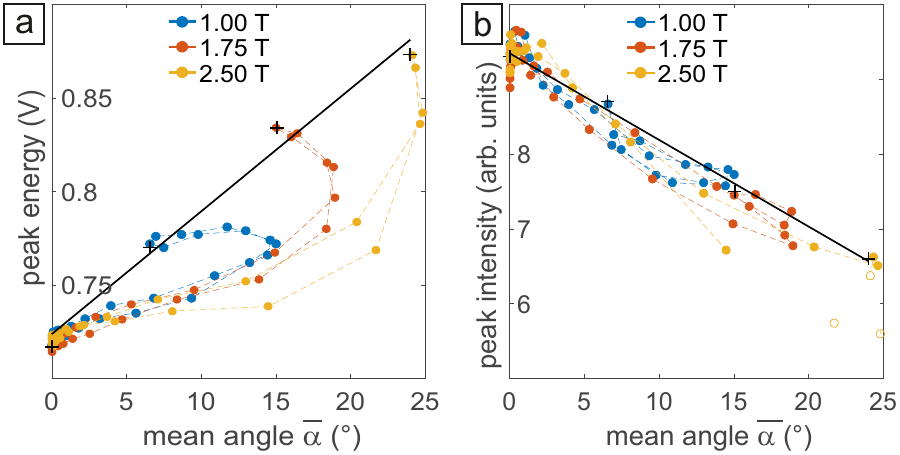}
\caption{\textbf{(a)}~Evolution of the peak energy (data of Fig.\,\ref{fig3}) as a function of mean angle at the respective position in the skyrmion, lines are guides to the eye. \textbf{(b)}~Same as (a) but for the peak intensity; the empty circles at the bottom mark data points that are considered unreliable due to a defect.}
\label{fig4}
\end{figure}

In our previous publication~\cite{Hanneken_NatNano2015} it was demonstrated that the peak energy in the center of a skyrmion scales roughly linearly with the external magnetic field and thus with the nearest-neighbor angle $\alpha_c$, see also inset to Fig.\,\ref{fig2}(e); the same was found for the full tight-binding calculations with small deviations at small angles~\cite{Hanneken_NatNano2015}. It is natural to try to generalize this dependence for the mean angle $\overline{\alpha}$ when considering arbitrary positions within the skyrmion. To analyze the relation between the experimentally observed peak energy and intensity with the degree of non-collinearity, we plot these parameters not as a function of the distance from the skyrmion center, but as a function of $\overline{\alpha}$ at the respective position, see Fig.\,\ref{fig4}(a,b). The three data points for the center of the skyrmions at the different magnetic field values are indicated by crosses, and the black line represents a linear fit to these data points and the ferromagnetic reference. While the data points for the intensity (b) have the tendency to be below the line, we conclude that the peak intensity in first approximation scales with $\overline{\alpha}$, irrespective of the applied field and the detailed spin configuration. This might be linked to a previous theoretical finding reported in Ref.\,\onlinecite{Seemann_PhysRevLett.108.077201}, in which the leading order contribution to the relative MR change due to the rotating magnetization was found to be proportional to the inverse of the spin spiral wave length, i.e.\ proportional to the nearest neighbor angles of a spin spiral.

In contrast, there are strong systematic deviations from a linear behavior for the peak energy, see Fig.\,\ref{fig4}(a): except for the positions in the skyrmion center the peak energy is significantly lower than the fit line, and even nearly constant for mean angles between $5-15^{\circ}$ at $B=-1$\,T. For a change of the angle in the skyrmion center of $10^{\circ}$ we find a peak shift of about 60\,mV; however, we can find the same peak shift for different positions within a skyrmion that have the same mean angle, e.g.\ $\overline{\alpha}=15^{\circ}$ at $B=1.75$\,T (at the skyrmion center and about 1.6\,nm away from it, see also red dot in Fig.\,\ref{fig2}(e)).

\begin{figure}[]
\includegraphics[width=1\columnwidth]{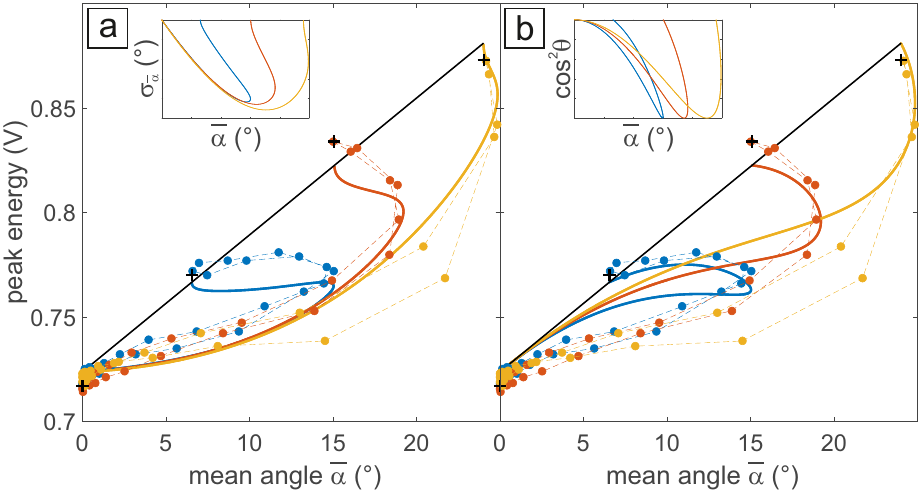}
\caption{Data as in Fig.\,\ref{fig4}(a) together with the evolution of \textbf{(a)} the degree of inhomogeneity $\sigma_{\overline{\alpha}}$ and \textbf{(b)} the quantization axis cos$^2(\theta)$ in addition to the mean angle; insets show the separate contributions of $\sigma_{\overline{\alpha}}$ and cos$^2(\theta)$ as a function of the mean angle.}
\label{fig5}
\end{figure}

We now come back to the details of the skyrmion spin texture, cf.\ Fig.\,\ref{fig2}(c): different positions can be characterized by the polar angle of the magnetization $\theta$, the mean angle $\overline{\alpha}$, the degree of inhomogeneity $\sigma_{\overline{\alpha}}$, and the quantization axis given by cos$^2(\theta)$. Because $\overline{\alpha}$ alone does not describe the evolution of the peak energy, we analyze whether one of the latter two parameters also plays a role: the insets to Fig.\,\ref{fig5}(a,b) show $\sigma_{\overline{\alpha}}$ and cos$^2(\theta)$ as a function of the mean angle, and the main panels demonstrate that either one can, together with the linear dependence on $\overline{\alpha}$, describe the general trend of the peak energy for all three magnetic field values. While the agreement using $\sigma_{\overline{\alpha}}$ seems to be better for positions closer to the ferromagnetic surrounding, the shape of the cos$^2(\theta)$ fits slightly better towards the center of the skyrmion.

\section{Conclusion}

We have studied the magnetic field dependent vacuum d$I$/d$U$ of skyrmions in the system of PdFe on Ir(111) and analyzed in more detail the previously observed peak, which is sensitive to the local spin texture, regarding the energy and the intensity. The peak intensity is found to be roughly linear with the local mean angle between neighboring magnetic moments, reflecting the degree of non-collinearity. In contrast, while the peak energy in the skyrmion center is proportional to its magnetic field dependent mean angle, we observe significant deviations from a linear behavior of the peak energy elsewhere in the skyrmion. We demonstrate that this could be attributed either to the inhomogeneity of the local spin texture as represented by the standard deviation of nearest neighbor spin angles, or to the mixing of states due to SOC, as manifested in TAMR. The latter depends on the local magnetization quantization axis and leads to a variation of the density of states from site to site across the skyrmion. While this has a direct effect on the band structure, also a more indirect effect is possible: when the bands that are altered by SOC are also involved in the NCMR, a slight modification of electronic states due to SOC may lead to a significant change of the density of states when the spin bands mix with each other, resulting in an enhanced NCMR signal. This possible interplay also implies that the relevant effects determining the peak shift need not be additive, as suggested by the simplistic illustration of Fig.\,\ref{fig5}, but can induce more complicated behavior of the LDOS.

The presented detailed experimental study can serve as a benchmark for future theoretical studies regarding the magnetoresistive properties of non-collinear states. Regardless of the details of the mechanism for the peak shift and intensity change the variation of the MR is large and we propose to use it as a means to detect and investigate the spin texture of non-collinear states.

\begin{acknowledgments}
We thank S.~Heinze, N.~Romming, and B.~Dup\'{e} for insightful discussions. Financial support from the German Research Foundation (DFG: GrK 1286 and SFB 668-A8) and the European Union (FET-Open project MAGicSky No.\ 665095) is gratefully acknowledged.
\end{acknowledgments}


\end{document}